# Spatially-resolved voltage-reversal due to Bernoulli potentials in dissipative $Bi_2Sr_2CaCu_2O_{8+x}$

*Sharadh Jois[1], Gregory M. Stephen[1], Samuel W. LaGasse[1], Genda Gu[2], Aubrey T. Hanbicki[1],

*Adam L. Friedman[1]

1   Laboratory for Physical Sciences, 8050 Greenmead Drive, College Park, MD 20740, U.S.A.

2   Condensed Matter Physics and Materials Science Department, Brookhaven National Laboratory, Upton, New York 11973, USA

*Corresponding authors: sjois@lps.umd.edu, afriedman@lps.umd.edu

## Abstract

We measure magneto-transport and critical currents in $Bi_2Sr_2CaCu_2O_{8+x}$ Hall bar devices. Above critical current in an applied magnetic field, we observe longitudinal differential voltage along one edge comparable in magnitude but opposite in sign to the other edge. This phenomenon is unaffected by reversal of the applied field, and seems unique to devices with invasive voltage contacts. We attribute the source of this behavior to particle-hole symmetry breaking in moving vortices and the formation of opposite Bernoulli potentials due to opposing vortex velocities at the edges where the invasive contacts create hotspots for rapid vortex nucleation and flux flow. These results are fundamental to the composition and flow of dissipative currents in layered superconductors.

## Main text

Superconductors placed in a magnetic field exhibit the Meissner-Ochsenfeld effect, i.e., perfect diamagnetism because of persistent screening currents on the surface. Rhoderick and Wilson demonstrated that the Meissner current in superconducting thin films flows along the periphery by using Hall effect sensors[1,2]. Several subsequent theories[3,4] and magnetometry experiments[5,6] reproduced this behavior in type-II superconductors. The dynamics of Abrikosov vortex nucleation and motion emerged using SQUID-on-tip techniques for nanoscale magnetic imaging[7,8]. Spatial modulations at the edge of superconductors, such as invasive voltage contacts[9], can constrict supercurrent flow and nucleate hotspots for vortex accumulation[7,10]. Vortices can also experience Bernoulli and Magnus effects, however, despite the wealth of research into superconducting devices, concrete evidence of mesoscopic Bernoulli potentials[11–13] is mostly unexplored.

$Bi_2Sr_2CaCu_2O_{8+x}$ (BSCCO), a high-$T_c$ two-dimensional superconductor[14], has generated much research interest because of observations of sign-reversal Hall effect[15], superconducting diode effect[16,17], Josephson diode effect,[18–20] and Klein-Andreev resonant states in junctions with graphene[21,22]. Despite the vast amounts of transport data, to the best of our knowledge, simultaneous measurements of the longitudinal voltages along both edges of a BSCCO device have not been explicitly reported. Magnetoresistance and Hall effect studies above critical current can yield essential mesoscopic information about composition and flow of charge carriers. Here, we characterize bulk BSCCO Hall bar devices as a function of temperature, current, and magnetic field. Additionally, we study a multi-modal Hall bar device designed to have non-invasive, invasive, and collinear contact geometries with increasing degrees of 'invasiveness' into the channel. Surprisingly, when the device with invasive contacts is measured in the presence of an external magnetic field below $T_c$ and the supplied current exceeds the critical current, we observe the longitudinal voltage along one edge of the device is opposite in sign to the voltage on the other edge. We explain this negative resistance as arising from rebalancing of kinetic energy due to vortex motion catalyzed by invasive contacts, leading to Bernoulli potentials. We perform a complete set of control experiments on multiple devices with different contact geometries and eliminate the extraneous effects that could affect our observations.

Hall bar devices are fabricated on 50-70 nm thick exfoliated flakes of BSCCO. An optical microscope image of a typical device is shown in **Figure 1a**. In this device, the BSCCO is 69±3.2 nm thick. Ohmic contacts are achieved by etching the top few layers and depositing Ag (10 nm) and Au (50 nm)[23]. We label the contacts C1-C6, as shown in Fig 1(a), and refer to them as such throughout this study. We show a schematic of our measurement setup (**Figure 1 (b)**) for applying a longitudinal DC current ($I_{DC}$) between C1 and C4 and measuring the longitudinal voltages $V_{xx,1}$ (C2-C3) and $V_{xx,2}$ (C6-C5) along both edges of the device. Detailed fabrication and methods are in the Supplemental Information.

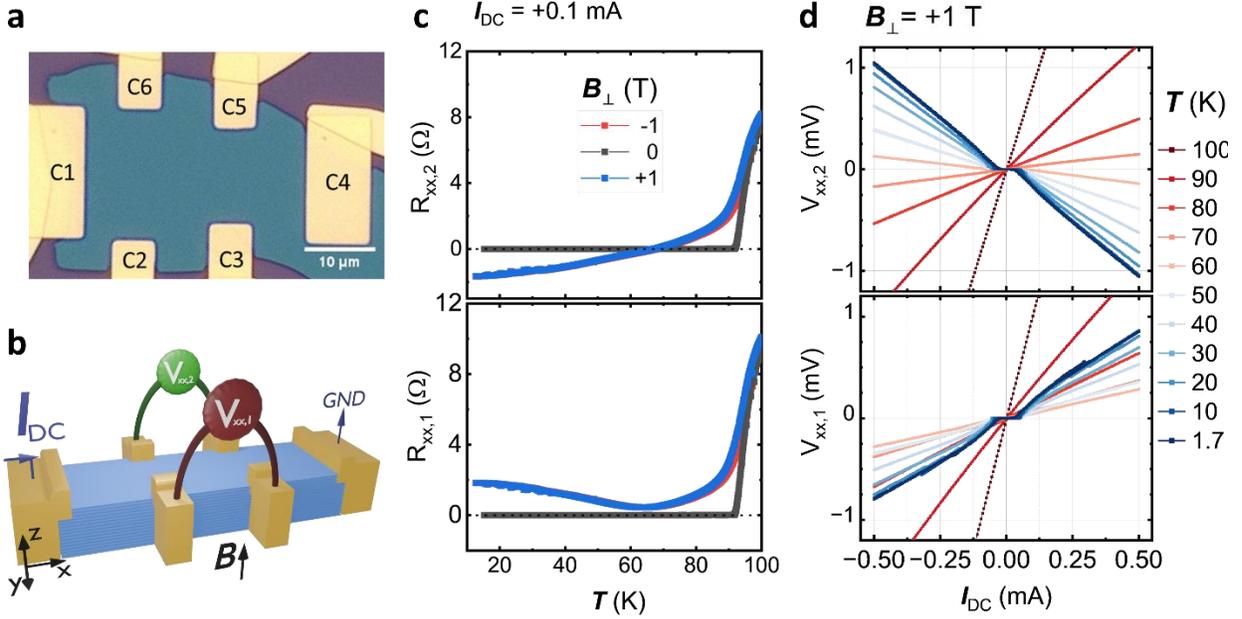

**Figure 1**. (a) Optical image of an example Hall bar device on BSCCO (69±3.2 nm). (b) Schematic of the electrical connections of a Hall bar device. We apply a DC current ($I_{DC}$) and measure the longitudinal voltages $V_{xx,1}$ (red) and $V_{xx,2}$ (green) along opposite edges of the device. (c) Temperature dependence of the longitudinal resistance ($R_{xx,1}$ and $R_{xx,2}$) monitored simultaneously with $I_{DC} = 0.1$ mA as the device is cooled in different external perpendicular magnetic fields $B_\perp = -1.0$ T (red trace), 0.0 T (black trace), and $+1.0$ T (blue trace). (d) *I-V* curves at different temperatures with a fixed perpendicular field of +1 T. Clearly $V_{xx,2} \approx V_{xx,1}$ in the normal state at 100 K (brown points) and $V_{xx,2} \approx -V_{xx,1}$ in the dissipative state below 60 K (blue points).

Longitudinal resistances on either edge of the Hall bar device ($R_{xx,1} = V_{xx,1}/I_{DC}$ and $R_{xx,2} = V_{xx,2}/I_{DC}$) are measured simultaneously. The temperature dependence of resistance while cooling the devices using a constant current bias of $I_{DC} = 0.1$ mA at zero applied magnetic field is the black trace in **Figure 1c**. Both $R_{xx,1}$ and $R_{xx,2}$ are comparable in value in the normal state, i.e., for T > 90 K, after which we observe a clear superconducting transition and a corresponding zero resistance characteristic of the superconducting state.

We then repeated the measurements with perpendicular magnetic fields of $B_\perp = +1.0$ T (**Fig. 1c** – blue trace) and $-1.0$ T (**Fig. 1c** – red trace). Note that it is difficult to distinguish the colors because they fall nearly on top of each other. The applied current, $I_{DC} = 0.1$ mA, is intentionally chosen to be slightly over the critical current ($I_c$) measured at $B_\perp = +1.0$ T when the sample is held at base temperature (1.7 K). These conditions create a measurable non-zero longitudinal voltage (resistance) in the device. As expected, because $I_{DC} > I_c$, the resistance decreases gradually between 90 to 70 K, but does not exhibit a superconducting transition because of the suppression of superconductivity in the Meissner state and vortex dynamics[15,24]. Below 60 K and when is $B_\perp = \pm 1.0$ T, the resistances on opposite sides of the device diverge $R_{xx,2} \approx -R_{xx,1}$ (**Figure 1c**). These opposing longitudinal resistances grow in magnitude with decreasing temperature. This observation was not mentioned in previous reports on sign-reversal of the Hall effect in Bi-based cuprate superconductors[15,25].

Further detail on this unique temperature-dependent resistance behavior is gleaned by examining the longitudinal voltages ($V_{xx,1}$ and $V_{xx,2}$) as a function of applied current. *I-V* curves with $I_{DC}$ varied from $-0.5$ to $+0.5$ mA at different temperatures in the Meissner state with $B_\perp = +1$ T, are shown in **Figure 1d**. In the normal state at T = 100 K, the voltages on opposite sides of the channel show Ohmic behavior with positive slopes. As the sample is cooled to below $T_c$, the slopes of both voltages gradually reduce until 70 K. Below 60 K, the slope of $V_{xx,2}$ abruptly changes sign and $V_{xx,2} \approx -V_{xx,1}$. Below 50 K, a clear superconducting region returns when $|I_{DC}| \leq |I_c|$, and both voltages are zero ($V_{xx,2} = V_{xx,1} = 0$). When $|I_{DC}| \geq |I_c|$, however, they again show opposite signs ($V_{xx,2} \approx -V_{xx,1}$). We verify that physically changing the current injection leads reverses the measured voltage (*supplemental Fig. S1*). The discrete voltage steps observed in the *I-V* curves between 10-20 K are phase slips[26,27]. The critical current in the Meissner state (*supplemental Figure S2*) peaks in this 10-20 K temperature region. The observation of high critical current and hysteresis between 10-20 K is attributed to the vortex glass transition[8,28].

The temperature dependent observations in **Figure 1** allow us to eliminate errors in sample wiring. Additional data confirming this sign reversal effect and ruling out trivial causes can be found in the Supplemental. *Supplemental Fig. S3* addresses the transverse ($V_{xy}$) component which also shows the sign reversal due to vortex dynamics[15,25] in the dissipative state. *Supplemental Fig. S4* shows data up to $\pm 12$ T from another device (57$\pm$1.4 nm thickness).

Colormaps of both longitudinal voltages $V_{xx1}$ and $V_{xx2}$ (**Figure 2a**) measured at base temperature ($T = 1.7$ K) show the sign of the voltages does not reverse with perpendicular magnetic field ($B_\perp$), but reverses only as a function of DC current. Positive voltage values are in shades of red, negative values are in shades of blue, and zero values in white corresponding to the superconducting regions. Comparing the two panels in **Figure 2a** beyond the superconducting regions, we immediately notice $V_{xx,2} \approx -V_{xx,1}$ and their signs reverse with current direction. For a fixed dissipative current $I_{DC} = +0.1$ mA, indicated by the dashed line in **Figure 2a**, there is an invariance in the sign of the measured voltages as a function of $B_\perp$ at different temperatures, as can be seen in **Figure 2b**. Here again, the voltages show consistent behavior ($V_{xx,2} \approx -V_{xx,1}$) below 40 K. The phase slips seen between 10-20 K in critical current measurements (**Figure 1**) are also evident in **Figure 2b** while sweeping magnetic field, manifesting as voltage steps. As the metallic state is reached at 60 K, we notice $V_{xx,2} \neq -V_{xx,1}$, consistent with resistance data in **Figure 1**. When the magnetic field is applied in-plane at an angle of 45° (*supplemental Fig. S5*) with respect to the transport current, we have weaker Josephson vortices[29] between the 2D individual layers and the superconducting region is broader due to the anisotropy of superconductivity in BSCCO. In normal metals, longitudinal voltages do not reverse sign with magnetic field. For BSCCO in the mixed state, a transport current has moving charges and vortices, therefore, one can expect the electric fields generated by vortex motion will create potentials which reverse with magnetic field. Our DC measurements clearly show that dissipative voltages do not reverse with field, like metals. However, the opposite sign of voltages measured at the top and bottom edges measured here would appear to violate the charge balance seen in conventional conductors. The Hall analysis below further expands on this notion.

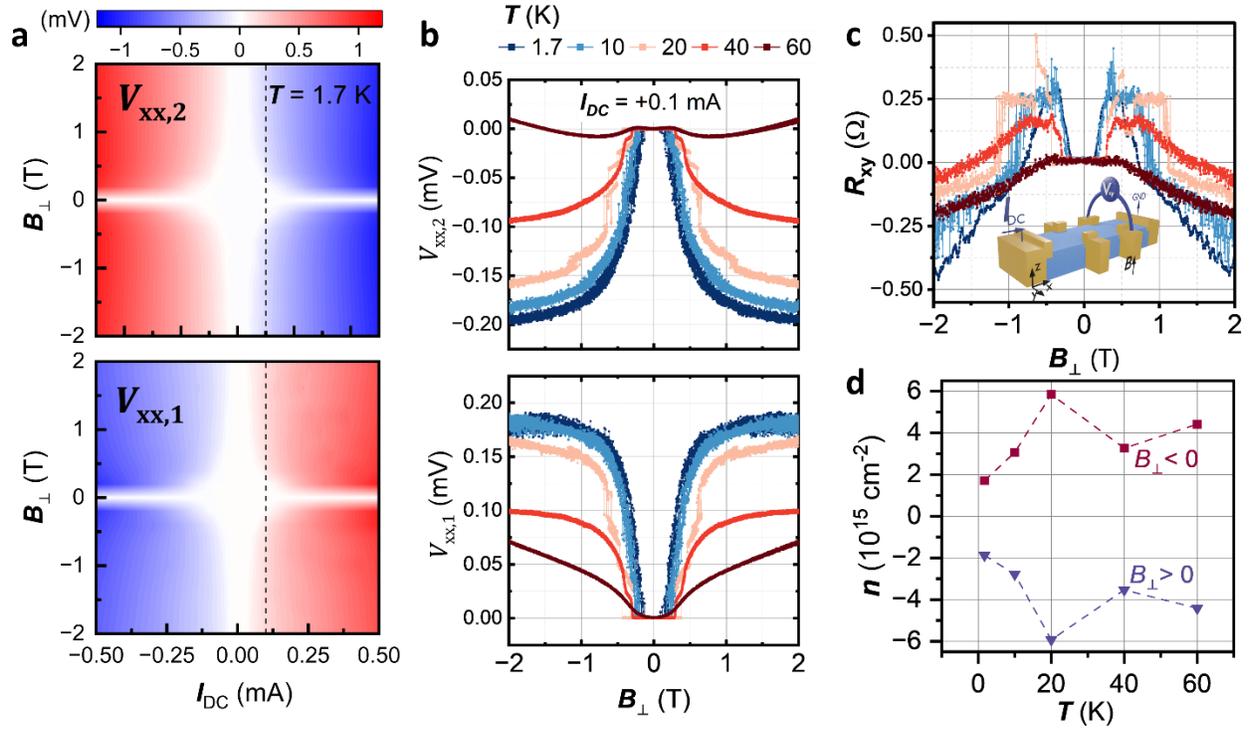

**Figure 2.** (a) Color plots of the longitudinal voltages $V_{xx,1}$ (bottom) and $V_{xx,2}$ (top) measured as a function of sweeping the current from -0.5 mA to +0.5 mA and stepping the perpendicular magnetic field ($B_\perp$) from -2.0 to +2.0 T in steps of 0.2 T. (b) The longitudinal voltages are shown as a function of magnetic field at different temperatures while sending a fixed current of $I_{DC} = +0.1$ mA. (c) Hall resistance (**not** anti-symmetrized) at different temperatures below 60 K. Inset shows a schematic for the Hall voltage measurement. The data are smoothed using a local regression (LOESS) to reduce the noise for Hall analysis. The negative linear regions of the data corresponding to $|B_\perp| \approx 2$ to 1 T are used estimate the carrier density in (d). The abrupt jumps to positive values are attributed to the random shape of the flake resulting in longitudinal components between $|B_\perp| \approx$ 0.5 to 1 T in (b) leaking into the transverse component. These jumps are not observed if the device has rectangular geometry (*supplemental Fig S8*). (d) The carrier density as a function of temperature shows hole-like charge when $B_\perp < 0$ (blue triangles), and electron-like charge when $B_\perp > 0$ (red squares). The lines are a guide for the eye.

The transverse voltage ($V_{xy}$) measured in the presence of a magnetic field corresponds to the Hall voltage. The sheet carrier density ($n_{sh}$) is related to the Hall resistance ($R_H = V_{xy}/I_{DC}$) by $n_{sh} = \frac{1}{q} \frac{1}{\left(\frac{dR_H}{dB_\perp}\right)}$ where $q$ is the charge constant. The sign of $\frac{dR_H}{dB_\perp}$ is indicative of hole (positive slope) or electron (negative slope) type carriers. Particle-hole symmetry breaking[30] appears in pseudogap systems, and above critical current there are different slopes, corresponding to hole-like and electron-like carriers, when the sign of the magnetic field is changed. The temperature and field dependence of the Hall resistance is shown in **Figure 2c.** For $T < 60$ K, $R_H$ is zero below a dissipative critical field ($B_{c1}$), and then jumps to a positive value. The sudden jump from zero to positive Hall resistance values in the low field regime seen here is due to the random shape of the BSCCO flake causing longitudinal components to bleed in as the sample transitions out of the superconducting state. Above 1 T, when the longitudinal components saturate, we obtain a negative Hall resistance due to vortex flux flow.

Analysis of the Hall data (**Figure 2d**) shows hole-type carriers with negative applied field (B < 0), and electron-type carriers with positive applied field (B > 0). We can understand this behavior based on the composition of charge carriers in the dissipative state below $T_c$ in the presence of a magnetic field. When excited above the critical current, Cooper pairs can be broken into single particles called Bogoliubov quasiparticles[31,32], which are superpositions of electron and hole wavefunctions[31] and obey particle-hole symmetry. The measurement of different carrier types by changing magnetic field is evidence for breaking particle-hole symmetry enabled by moving vortices[33] along the transverse direction. The magnitude of the carrier densities of broken Cooper pairs ($n_s$) in the dissipative mixed state (B > $B_{c1}$) decreases from $5 \times 10^{15}$ cm$^{-2}$ at 60 K to $2 \times 10^{15}$ cm$^{-2}$ at 1.7 K, and shows a spike at 20 K when there is a vortex glass transition[8]. In comparison, the Hall analysis in the normal state at 100 K (*supplemental Figure S6*) shows only hole-type normal-state carrier density $n_N = +18.85 \times 10^{15}$ cm$^{-2}$ and mobility $\mu = 14.7$ cm$^2$/s. The ratio of superconducting carriers to normal state carriers ($n_r = n_s/n_N \approx 10\%$) is reasonable and is used later in the calculation of Bernoulli potentials.

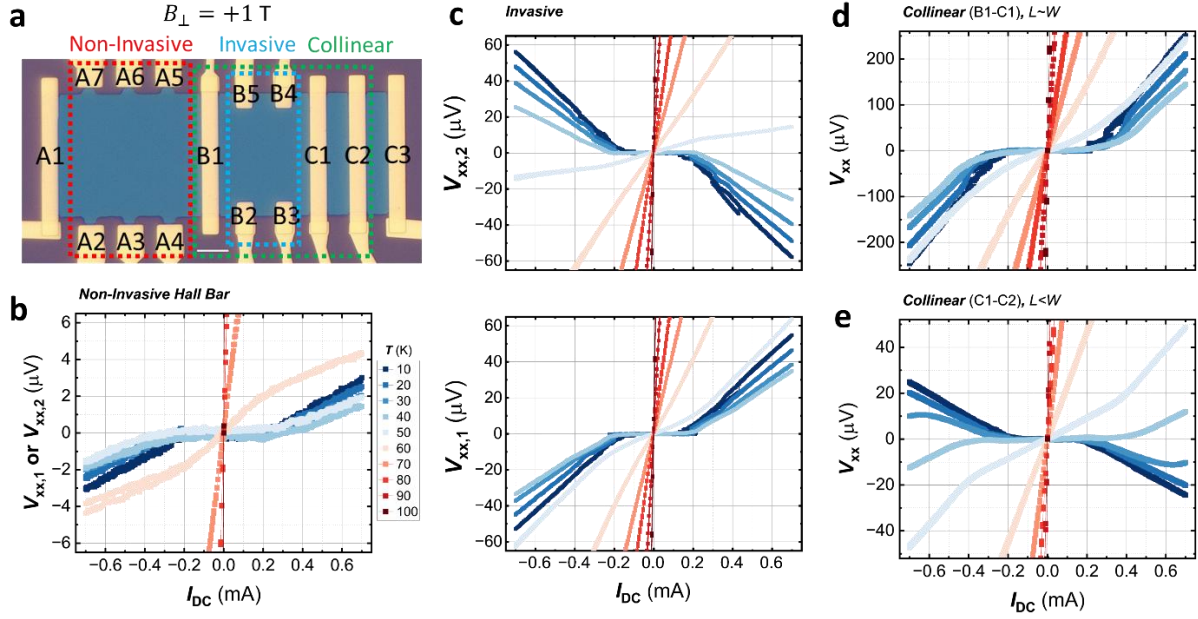

**Figure 3**. (a) Optical image of the "3-in-1" device with non-invasive, invasive, and collinear contacts. The white scale bar below B1 corresponds to 10 µm. The width of the channel is $W = 40$ µm and thickness 72±2.9 nm. We show data at $B_\perp = +1$ T and verify similar behavior in negative field. (b) Voltages measured on both sides of the channel with the non-invasive contacts at all temperatures. The temperature legend in this plot also applies to all data panels of this figure. The small signals measured here are due to the dissipative normal state. (c) Longitudinal voltages, $V_{xx,2}$ (c, top, B5-B4) and $V_{xx,1}$ (c, bottom, B2-B3) from the invasive part of the device shown in (a). Note, this region resembles the Hall bar devices discussed earlier in Fig.1. (d) The collinear voltage measured across the invasive region, B1-C1 (L~W). (e) The collinear voltage across the short channel, C1-C2 (L<W).

The geometry of the voltage contacts is a crucial component to the observation of opposite voltages across the channel. In particular, it has been reported in thin Pb films that constrictions can create local hot spots for rapid vortex nucleation and bunched transport[7]. Furthermore, in quantum transport[9], voltage contacts encroaching into a conducting channel perturb current flow and cause invasive effects[34]. To explore such effects, we fabricated a BSCCO device with three different contact geometries ("3-in-1 device")– non-invasive, invasive, and collinear, as shown in **Figure 3a**. The thickness of the 3-in-1 device is 72±2.9 nm. The non-invasive region has small tab-like extensions of the superconducting material protruding away from the channel where the contacts are made. The region with invasive contacts is like the device in **Figure 1**, a standard Hall bar geometry made on exfoliated flakes of 2D materials. The collinear contacts spread across the entire width of the device and emulate the extreme case of invasiveness. A superconducting transition ($T_c = 87$ K) is seen in all parts of the device. In the experiment, we inject DC current at A1, ground C3, and measure longitudinal differential DC voltages at various combinations of intermediate contacts. Above 50 K (light blue), all voltages in **Figure 3 b-e** show positive slopes as a function of current. The voltages measured with the non-invasive contacts above $I_c = 0.2$ mA on both sides of the channel, **Figure 3b**, are an order of magnitude lower than the invasive contacts ($< 4$ µV) and always show positive slopes, i.e., $V_{xx,2} = V_{xx,1}$. The invasive contacts, **Figure 3c**, show $V_{xx,2} \approx -V_{xx,1}$ below 50 K, consistent with behavior seen in the irregularly shaped BSCCO flakes reported earlier**Error! Reference source not found.**. The observation in the invasive region of the device is supported by additional results from single ended voltage experiments (*supplemental Figure S7*) and series resistance analysis (*supplemental Figure S8*) in the invasive region. Negative Hall resistance is seen in the invasive region, but not in the non-invasive region of the 3-in 1 device (*supplemental Fig. S9*).

The collinear contacts are an extreme version of invasiveness and in this device we measure two cases in **Figure 3d** with $L \sim W$, (B1-C1) and **Figure 3e** with $L < W$, (C1-C2). In the first case, L~W, the *I-V* curves are always positively sloped. In the second case, however, below 30 K the slope transitions from positive ($0.7 < |I_{DC}| < 0.6$ mA) to negative when $|I_{DC}| < 0.6$ mA. The length between contacts C1 and C2 is comparable to the length between the invasive voltage contacts B2 and B3 (or B5 and B4). The opposite resistance behavior seen in the top invasive contacts (B5-B4) and short channel collinear contacts (C1-C2) strongly suggests the mechanism for negative resistance is related to vortex confinement and edge effects.

A plausible mechanism for the observed voltage-reversal effect is inspired by the Bernoulli effect in fluid dynamics where kinetic and potential energies balance out. Here, we present a version of the Bernoulli effect caused by moving charge carriers and vortices. When BSCCO in the mixed state is driven into the dissipative regime by applying large currents, the rebalancing of kinetic and potential energies results in local Bernoulli potentials ($\Phi_B$) [11,12,35]. In equilibrium, the Bernoulli potential is,

$$\Phi_B(x,y) = -n_r \frac{1}{2q} m^* |\Sigma v(x,y)|^2 \qquad \text{Eq. 1}$$

Here, $q$ is the charge constant, $n_r$ is the ratio of superconducting density to the normal state carrier density as described above, $m^* = 4m_e$ is the effective mass of carriers[36] and $\Sigma \boldsymbol{v}(x, y)$ is the sum of velocities from carrier transport ($\boldsymbol{v}_{transport}$) and vortex motion ($\boldsymbol{v}_{vortex}$).

To explain our proposed mechanism, we have generated a detailed schematic of a BSCCO device with the various particles, forces, parameters, and coordinate systems indicated (**Figure 4a**). We will refer to this figure in the following discussion. The device has a pair of invasive voltage contacts (gold) along the top and bottom edges. A transport current ($\boldsymbol{J}$) is sent along $+\hat{x}$. It is known that constrictions[7] and periodic indentations[10,37,38] can amplify vortex bunching and rapid transport. The invasive contacts in our BSCCO devices create edge indentations and will therefore increase current crowding. The surrounding areas then become hotspots for vortex nucleation. Above critical current, the transport velocity of carriers is dominated by the gap velocity ($v_{gap} = 1.76 \times 10^4$) from breaking Cooper pairs[39]. An external perpendicular magnetic field ($\boldsymbol{B}$) along $+\hat{z}$ produces anti-clockwise vortices (mixed state) in the BSCCO that move when the Lorentz force ($\boldsymbol{F}_L = \boldsymbol{J} \times \phi_0 \hat{z}$) exceeds the pinning force. Here, $\phi_o$ is the magnetic flux through a single vortex. Vortices move towards $-\hat{y}$ with Lorentz velocity $\boldsymbol{v}_L$, overcome viscous drag,[40] and experience a Magnus force leading to a longitudinal electric field $\boldsymbol{E} = -\boldsymbol{v}_L \times \boldsymbol{B}$. Note that transverse vortex motion has been recently been imaged by SQUID-on-tip techniques[7]. Reversing the magnetic field also reverses the Lorentz force. Therefore, the electric field is not changed, explaining why the sign of measured voltages does not change with reversing field direction.

Vortices accumulate towards the bottom edge and collectively have a circulation velocity. At the top edge, the transport velocity and vortex circulation are in opposite directions and partially cancels, $\boldsymbol{v}_{top} \approx \boldsymbol{v}_{transport} - \boldsymbol{v}_{vortex}$. The vortex circulation at the bottom edge is along the transport current and adds, $\boldsymbol{v}_{bot} \approx \boldsymbol{v}_{transport} + \boldsymbol{v}_{vortex}$. The reduced velocity at the top edge results in a net positive Bernoulli potential and vice versa at the bottom edge. Note, to represent the modulations to the vortex velocities created by the contacts, we introduce a sinusoidal factor. The electric field corresponding to the Bernoulli potentials is $\boldsymbol{E} = -\Delta \Phi_B$. Detailed descriptions of the forces and velocities used to then simulate the Bernoulli potential are in the *supplemental material*, *Fig. S10*.

Substituting the velocity terms from dissipative transport and modulated vortex motion around the contacts (supplement) into **Eq. 1**, we quantitatively map the Bernoulli potentials, **Figure 4(b)**. The invasive, rectangular contacts (yellow) have an overlap width of 5 µm into the channel. Their sharp Gaussian potentials are included in the calculation for completeness. The calculation shows bunched potentials appearing at the invasive contacts on the right at +5 µm. From this map, we note the differential voltage along the bottom is $V_{xx,1} = 68.0$ µV and $V_{xx,2} = -68.7$ µV and closely reproduces the experimental finding ($V_{xx,2} \approx -V_{xx,1}$).

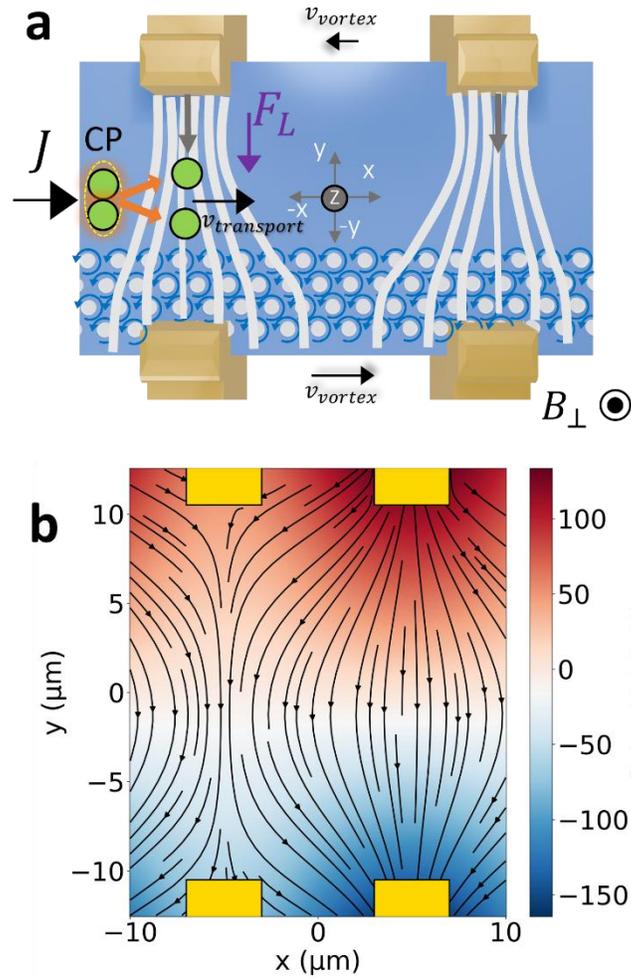

**Figure 4.** (a) The schematic shows vortices accumulating along the bottom edge when a current ($J$) is injected along +x in a BSCCO Hall bar device with invasive contacts placed in an external perpendicular magnetic field ($B_\perp$). Above critical current, the Cooper pairs (CP) separate into unpaired quasiparticles traveling at the gap velocity. Vortices rapidly nucleate around the invasive contacts. A Lorentz force $F_L$ pushes them down towards -y. At high currents and magnetic fields, the transverse vortex flow appears

like streams (white curved lines) to the bottom contacts. The ensemble of vortices piling up at bottom have a circulation velocity that adds to the net transport velocity at the top (subtracts at the bottom) and create the Bernoulli potential. (b) Using the model described in the main text, the calculated two-dimensional potential $\Phi_{total} = \Phi_B + \Phi_{contacts}$ and electric field profile $E = -\Delta\Phi_{total}$ for a device (length=20 µm, width = 25 µm, thickness = 50 nm) with similar dimensions to experiment. The separation between contacts is 10 µm.

The model described here is a simplified vector model to show Bernoulli potentials can arise in a BSCCO Hall bar device with invasive contacts. In contrast, we expect the modulation to the circulation velocity of the vortex ensemble to be negligible when the contacts are non-invasive. This would explain why we do not see negative longitudinal voltages in experiment for non-invasive Hall bar devices. Lastly, short channel collinear contacts create boundaries where vortices are trapped. This can create a large circulating current of the vortex ensemble leading to an overall negative Bernoulli potential. Detailed finite-element analysis paired with time-dependent Ginzburg-Landau equations are required to completely model all the non-equilibrium transport dynamics observed here.

The superconducting diode effect[16] and Josephson diode effect[18,20] are pivotal experiments showcasing the non-reciprocal transport with respect to changing current direction[17] in BSCCO. Our work adds to this canon and displays spatially resolved opposite edge resistances in BSCCO devices due to opposite Bernoulli potentials and particle-hole symmetry breaking. Invasive voltage contacts separated by short distances (< 10 µm) create ratchet potentials and confinement[10,37]. Regions surrounding the invasive contact become hot-spots for nucleating vortices. The motion of vortices relative to dissipative carriers causes rebalancing of potentials near the contacts, resulting in negative resistance along one edge. Future work on exfoliated BSCCO and other layered superconductors should consider these effects by measuring the longitudinal voltages along both sides of Hall bar devices. Optionally, using non-invasive voltage contacts may alleviate negative resistances from vortex-flow Bernoulli potentials.

## Conclusion

In this work, we characterized Hall bar devices of BSCCO and found the longitudinal voltages measured on either side of the Hall bar have opposite signs ($V_{xx,2} \cong -V_{xx,1}$) when the supplied current exceeds the critical current, i.e., $I > I_c$. The results are reproducible in multiple devices of exfoliated BSCCO flakes (50 to 70 nm thickness). We propose that vortex flow across invasive contacts induces a Bernoulli potential that creates the observed opposite longitudinal voltages. The effect is observed in out-of-plane and in-plane magnetic fields, suggesting both pancake vortices and Josephson vortices can cause this effect. Additionally, through Hall measurements we detect charge carriers changing sign with magnetic field as evidence for particle-hole symmetry breaking, likely in pancake vortices. The observations emphasize the role of invasive contacts becoming nucleation centers for bunched vortex motion in devices of unconventional superconductors[41]. This effect can be harnessed to create negative resistors and low-power voltage inverters for superconducting logic where naturally asynchronous $\frac{\pi}{2}$ sources may be needed.

**Author Contributions**

G.G. grew the BSCCO crystals. S.J. noticed the effect, conceived the experiments, and fabricated the devices. S.J. measured the devices with assistance from G.S. A.F. and A.H. supervised the experiments. S.J. formulated the explanation and drafted the manuscript with input from A.F., A.H, and S.W.L. All authors contributed to the final version of the manuscript.

**Acknowledgements**

We thank the clean room staff at LPS – K. T. Kim for assistance in fabrication, C. Walsh and T. Kirlew for equipment support. We also acknowledge the contributions of colleagues Dr. N. Smith, Dr. V. Sharma, & Dr. S.T. Le for discussions and assistance in the lab.

We appreciate discussions with Prof. J. U. Lee, Dr. M. J. Balakan, Prof. S. Ulloa and M.A.M. Ramirez, and Prof. E. Zeldov.

**Funding Acknowledgement**

The work at Brookhaven was funded by the U.S. Department of Energy (DOE) the Office of Basic Energy Sciences, under Contract No. DE-SC0012704.

**Data Availability**

All relevant data is already presented in the manuscript. Raw data will be made available upon request to the corresponding authors.

# Supplementary Information

**Methods**

The optimal doped Bi-2212 single crystals with $T_c = 91$ K was grown by using a floating zone method[42]. BSCCO crystals are cleaved and exfoliated on 285 nm silicon dioxide wafers in an inert argon-filled glovebox (<0.1 ppm $O_2$ and $H_2O$). The wafers are pre-patterned with an alignment mark array to help with recording the position of target flakes. Flakes of interest are identified after scanning the entire chip under a microscope of the customized 2D Factory system and generating a stitched image. Candidate flakes of thickness ranging from 50 to 70 nm are selected for Hall bar devices, like the one shown in **Figure 1a**. The devices are fabricated in a cleanroom using direct-write laser lithography within the same day. First, the contact pads are made with Ti (10 nm) and Au (50 nm) around the target flake. Then, Ag (10 nm) and Au (50 nm) are used for Ohmic contacts to BSCCO. Additional information can be found in previous work [21–23].

The finished devices are tested for contact resistance. Then the wafers are cleaved, mounted on a 16-pin chip carrier and gently wedge bonded with Al wire. These are immediately loaded into an Oxford Teslatron ($T_{base} = 1.7$ K, $B_{max} = \pm 12$ Tesla) for measurement. The data is collected using a Nanonis Tramea system with two voltage pre-amplifiers (4 + 4 channels) with up to 1000X gain, capable of doing single-ended or differential AC lock-in or DC voltage measurements. A voltage driven SRS 580 current source (1 mA/V) is used to supply the DC current. The device is grounded to the SRS 570 current pre-amplifier which measures the current signal in the device to voltage (1 mA/V). All measurement electronics are grounded together. We verified the effect is reproducible in another cryostat ($T_{base} = 3$ Kelvin, $B_{max} = \pm 1$ Tesla) using Keithley 2182 nanovoltmeters.

Calculations were done with the help of Google Gemini Pro to construct the initial python script. The code was tested and corrected to represent experimental findings.

**Reversing Current direction**

For consistency, we check that the measured voltages reverse slope by changing direction of current bias in **Figure S 1**. The schematics show the circuit for +I$_{src}$ (**Figure S 1 (a)**) and -I$_{src}$ (**Figure S 1 (b)**). The corresponding data in **Figure S 1 (c)** shows the measured voltages $V_{xx,1} = -V_{xx,2}$ for sourcing +I$_{src}$ (blue). Reversing the injection of current to -I$_{src}$, reverses the slopes of these voltages (red).

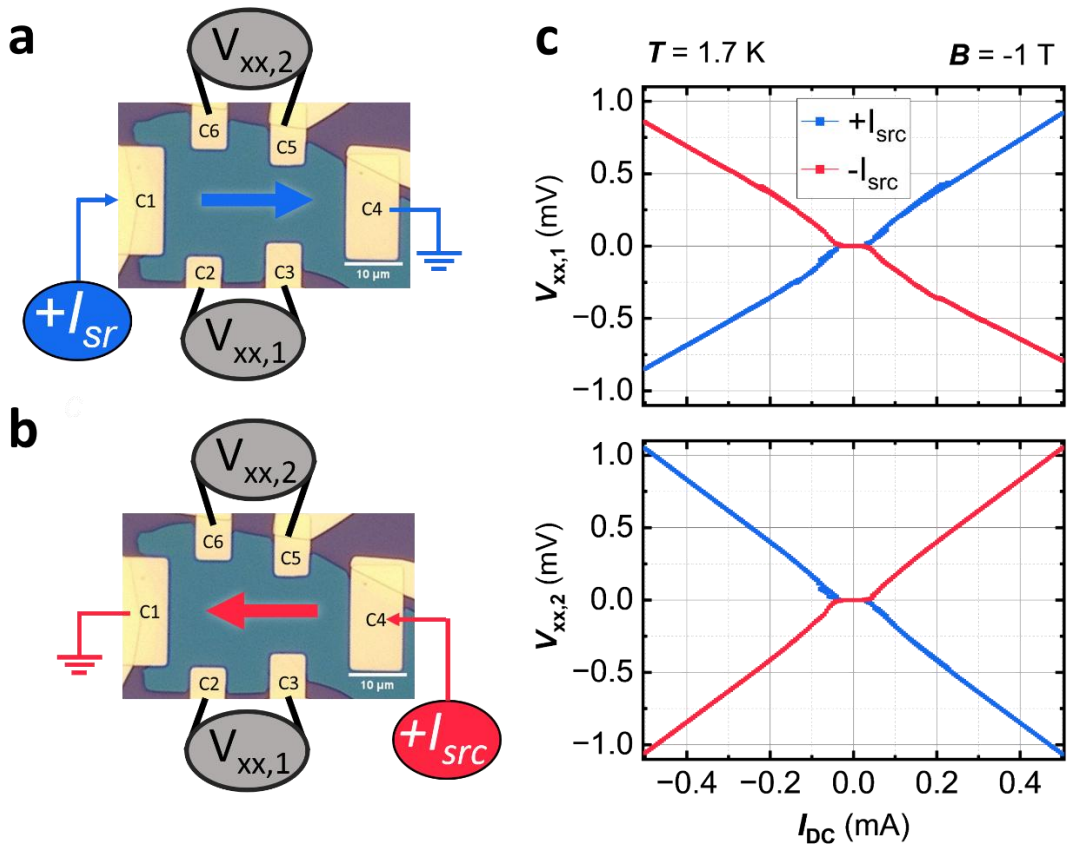

**Figure S 1.** Schematics of connections to source +I$_{src}$ (a) and -I$_{src}$ (b), and measure fixed longitudinal voltages $V_{xx,1}$ and $V_{xx,2}$ shown in (c). We observe the measured voltages above critical current change slope when current direction is reversed.

**Critical current below 50 K**

The critical currents from the forward sweep (-0.5 to +0.5 mA) and reverse sweep (+0.5 to -0.5 mA) are extracted in the Meissner state for data shown in **Figure 1 (d)** of the main text. Here we reproduce the data below 50 K in **Figure S 2 (a)**. From this data we extract the positive critical current ($+I_c$) from the forward sweep and the absolute value of the negative critical current ($|-I_c|$) from the reverse sweep and show it in **(b)**. Taking the reading of the current when the DC differential voltage reading is $< 0.4$ µV is criteria for determining the critical current. Interestingly, the critical current in the Meissner state peaks between 10 to 20 K, the same temperature range when the phase slips (jumps in voltage) are the largest. Lower critical currents are seen at the base temperature of 1.7 K in the Meissner state.

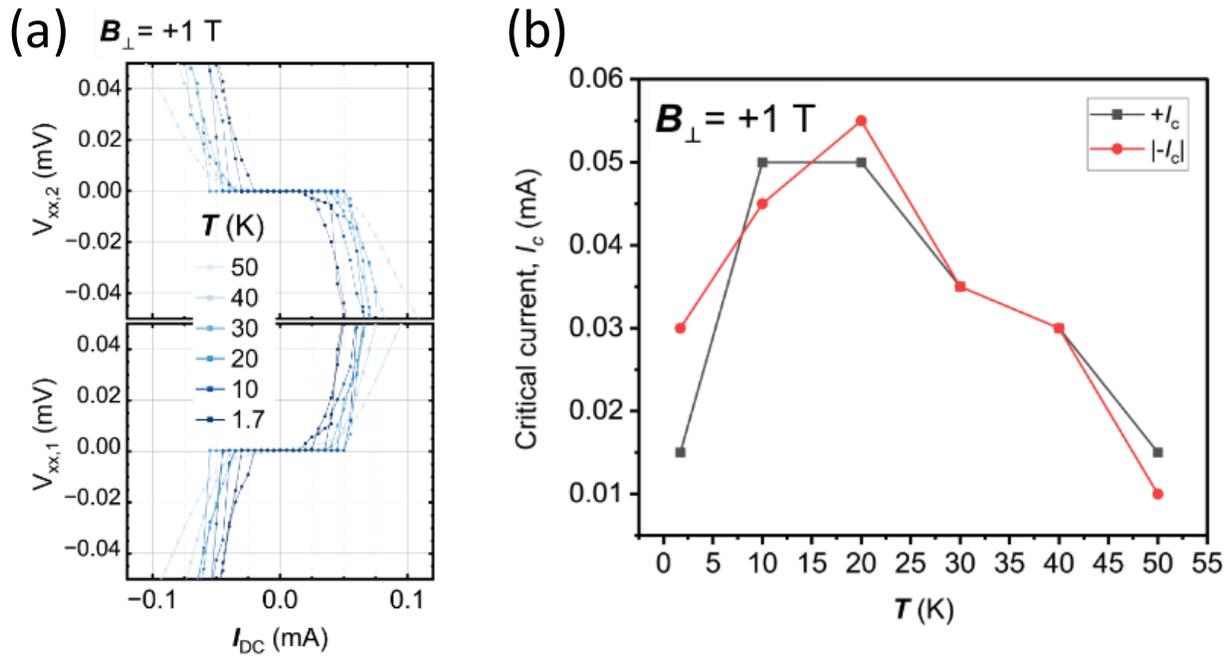

**Figure S 2**. a) IV in the low current regime for temperatures 50 K and below, and b) critical current based on these measurements.

**Transverse resistance & voltage as a function of temperature**

The transverse resistance measured as a function of temperature at different magnetic fields is shown in **Figure S 3 (a)**. Here we see the sign reversal of the Hall resistance at $\pm 1$ T first at $T_{c,B=0} = 90$ K, and then again at 60 K. This region was previously studied in detail [1] and attributed to vortex dynamics. The difference between -1.0 T data and +1.0 T data between 70 to 90 K is due to changes in the natural rate of cooling of the cryostat. The abrupt jump in resistance at 20 K is due to the vortex glass transition [2]. In **Figure S 3 (b)**, the transverse voltage vs. $I_{DC}$ sweep taken at different temperatures is shown. Here, we see the slope is positive in the normal state between 100 and 90 K and the slope changes to negative between 80 to 60 K. Below 40 K, we see a positive slope when $I_{DC} \geq |I_c| = 0.05$ mA to 0.1 mA. When $|I_{DC}|$ exceeds 0.1 mA negative slope is seen again. At 10 and 20 K, we see many steps due to phase slips in the critical current like we see in the longitudinal voltages. The sign reversal of the slope behavior seen in the critical current measurements of the transverse voltage are consistent with $V_{xx,2}$ discussed in the main text. This supports the presence of a transverse electric field gradient above critical current. The boxed region at low current is enlarged and shown in **Figure S 3 (c)** at a few temperatures to emphasize the mix of positive and negative slopes appearing in the transverse voltage. This sort of anomaly is attributed to invasive contacts on randomly shaped BSCCO flake device.

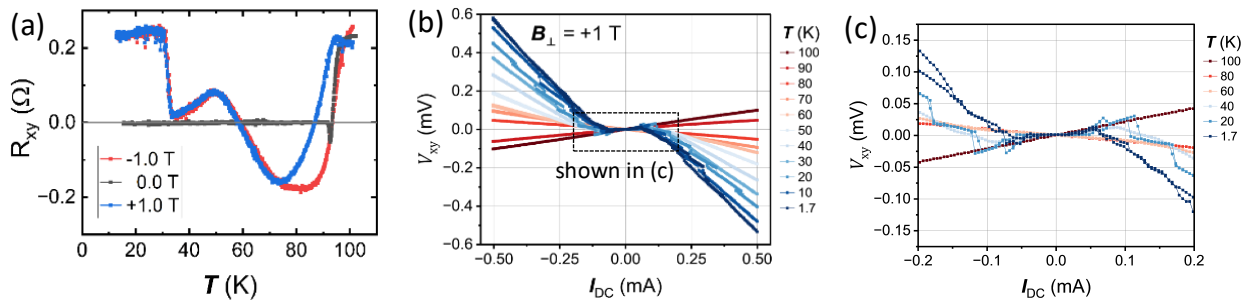

**Figure S 3**. (a) Transverse Resistance vs. Temperature at different magnetic fields. (b) Transverse voltage as a function of sweeping $I_{DC}$. These data go along with the longitudinal data of Figure 1 in the main text. The boxed region is enlarged and shown in (c) at a few temperatures.

**Data from another BSCCO device**

We support the reproducibility of our results by showing data from another BSCCO device in which we observed the same behavior. In this device, the BSCCO is 57±1.4 nm thick. In **Figure S 4** (a) and (b) below, we again see $V_{xx,2} \cong -V_{xx,1}$ even at high magnetic fields up to ±12 T. Both longitudinal voltages regain conventional Ohmic behavior above 60 K, as shown in (c) and (d).

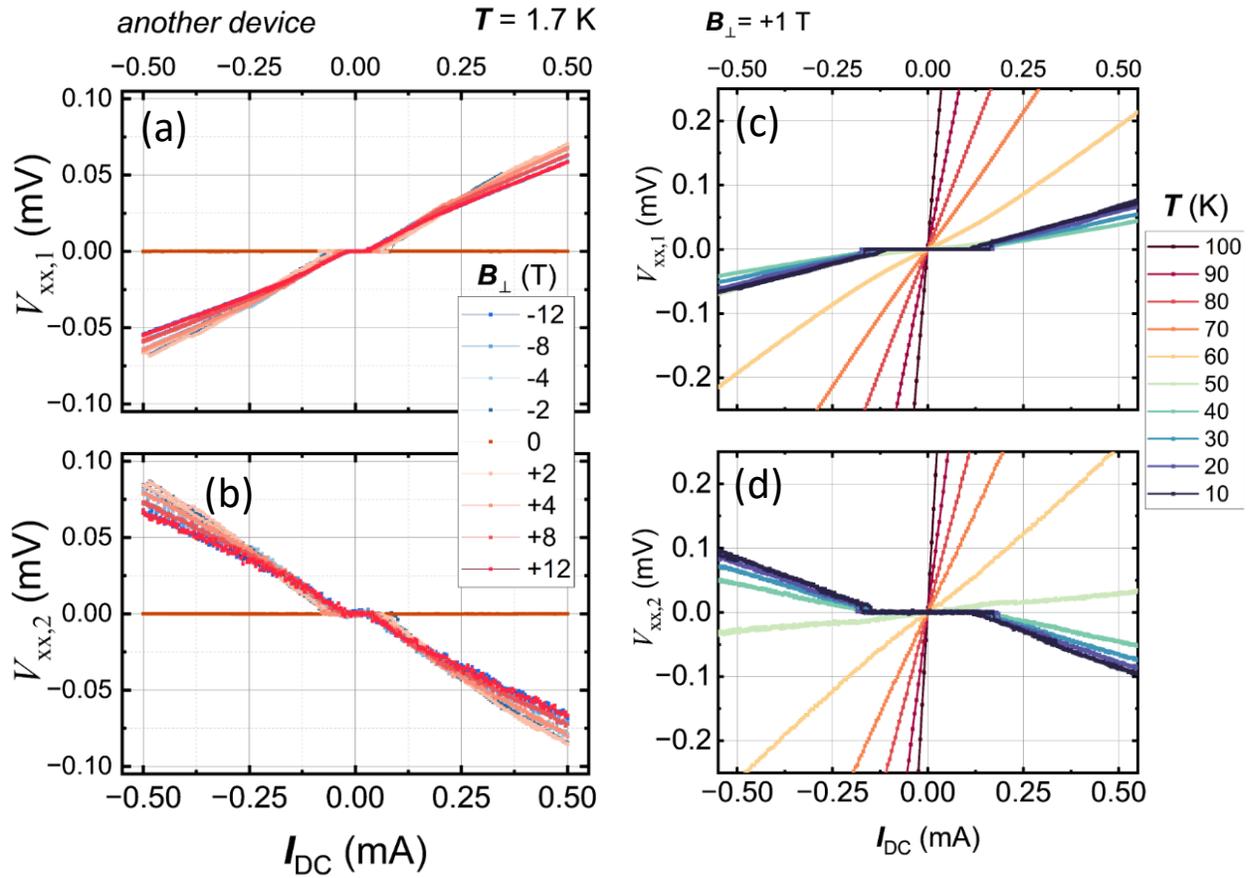

**Figure S 4**. Panels (a) and (b) show the longitudinal voltages measured in another BSCCO device at base temperature T = 1.7 K and different magnetic fields. The same components are shown at fixed field of +1.0 T at different temperatures in (c) and (d). This figure supports the reproducibility of the behavior described in this work.

**In plane magnetic field**

To support to the out-of-plane magnetic field data in the main text, in **Figure S 5** we show the voltages do not change sign as a function of in-plane magnetic field ($B_\parallel$) applied at an angle of 45 ° with respect to the supplied current. The longitudinal and transverse voltages are shown in **Figure S 5 (a-b)** and **(c)** respectively.

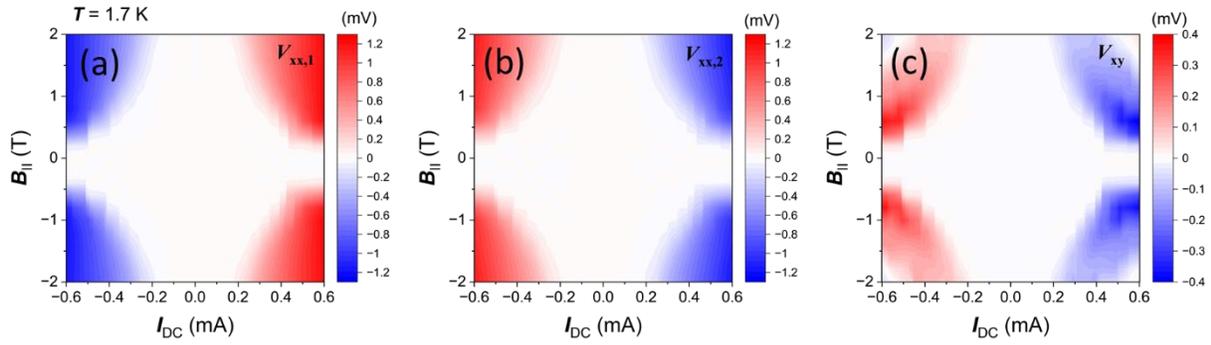

**Figure S 5**. Color plots of the longitudinal voltages $V_{xx,1}$ in (a) and $V_{xx,2}$ (b), and transverse voltage $V_{xy}$ in (c) measured as a function of sweeping the current from -0.5 mA to +0.5 mA and stepping the in-plane magnetic field ($B_\parallel$) from -2.0 to +2.0 T.

**Hall Resistance in the normal state at 100 K**

Measurements above T_c show our sample has a metallic behavior, **Figure S 6**. The Hall resistance (anti-symmetrized) of a BSCCO device measured at 100 K is shown below. The fit using the Hall equations yield a carrier density ($n_{sh}$) and mobility ($\mu$), given in the inset.

$$n_{sh} = \frac{1}{q_e \frac{dR_{Hall}}{dB_\perp}} \quad \text{Eq. S1}$$

$$\mu = \frac{1}{R_{sheet}} \times \frac{dR_{Hall}}{dB} \quad \text{Eq. S2}$$

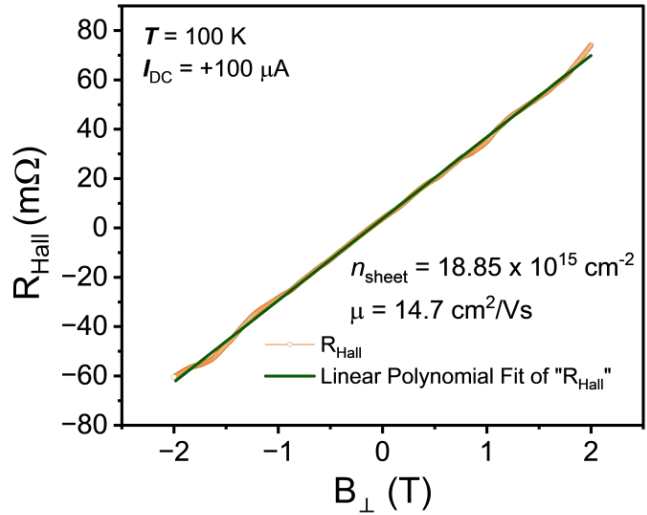

**Figure S 6**. Anti-symmetrized Hall resistance of a BSSCO device measured at 100 K showing conventional hole-type carriers at 100 K. The dark green line is the fit to extract sheet carrier density ($n_{sheet}$) and mobility ($\mu$) shown in the annotation.

**Additional Characterization of the 3-in-1 Device**

1. Single-ended voltage

In this section we show the mathematical difference of single-ended voltages shows the same observation of opposite signs above the critical current as the differential voltages presented in the main text. In **Figure S 7 (a)**, for completeness, we show the two-terminal voltage measured across the source and ground leads while sweeping the DC current from -0.7 to +0.7 mA. The single-ended voltages measured on the invasive Hall bar voltage contacts B2 to B5, **Figure S 7 (b),** are all nearly half the two-terminal voltage. These single-ended voltages were measured simultaneously using the Nanonis Tramea system and are measured with respect to a common ground. The current pre-amplifier converts the measured current signal to voltage (1mA/V), which is also measured by the Nanonis. This is because most of the voltage drop occurs at the source (A1) and ground (C3) contacts which have much larger resistance (~400 Ω) compared to the channel resistance (< 1 Ω). Nonetheless, the large precision of these single-ended voltages enables us to take the mathematical difference and verify the main result. The difference $V_{B2} - V_{B3}$ and $V_{B5} - V_{B4}$ shown in **Figure S 7 (c)** closely resemble the measured differential voltage in **Fig. 3 (c)** of the main text. This test eliminates contributions from multiple connections and wiring errors. This test also eliminates the possibility of artifacts from differential voltage amplifiers.

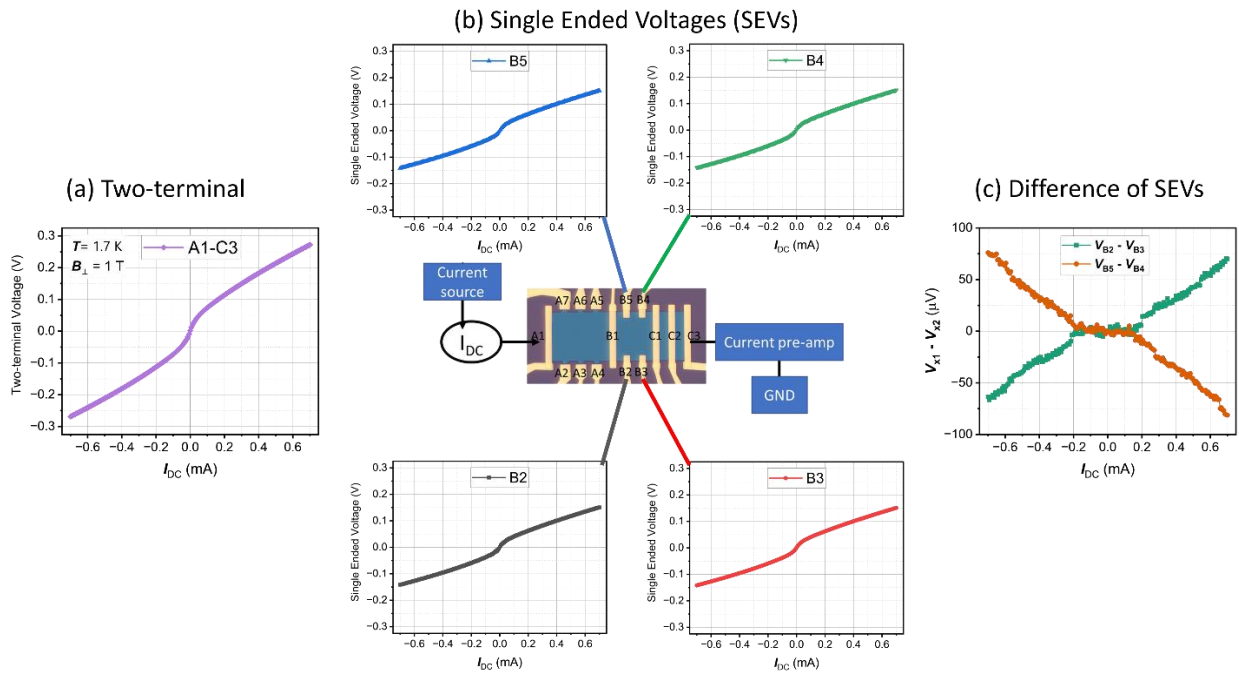

**Figure S 7. (a)** Two-terminal voltage measured across the source (A1) and ground (C3) contacts while also measuring the single-ended voltages on B2 to B5 for the device shown in **(b)** at base temperature of 1.7 K in a perpendicular magnetic field of +1 T. **(c)** The mathematical difference of single-ended voltages shown in **(b)** gives us the same observation discussed in the main text.

2. Series resistance analysis

Another control to support the result in the main text is treating the invasive Hall bar region as a series of two resistors under constant current. Knowing the voltage change over the first and second resistors, it becomes straightforward to calculate the voltage change between the first and second resistor, i.e., $V_{x1-x2} = V_{high-x2} - V_{high-x1}$. The experimental results for this hypothesis are discussed below.

The optical image of the device with current source at A1 and ground at C3 is in **Figure S 8 (a)**. We first measure the DC differential voltage B2-B3 and then measure B5-B4 while sweeping the DC current from -0.7 to +0.7 mA. Both are shown together in **Figure S 8 (b)**. This result is comparable to the low-temperature data in **Fig. 3 (c)** of the main text. Then, while keeping the current path the same, we connect the voltage-high to B1 and vary the voltage-low connection from B2 to B5, **Figure S 8 (c)**, and measure these DC differential voltages for the same current sweep in **Figure S 8 (b)**. While using a pair of voltage contacts, all other contacts were set to float. The first surprising result here is the slope in the normal region is already sign reversed with respect to B1, i.e., there is a large negative potential somewhere between B1 to C1 despite positively forced current. This negative potential region is the same Bernoulli potential described in the main text. Here we can already see $V_{B1-B2} > V_{B1-B3}$ and $V_{B1-B4} > V_{B1-B5}$.

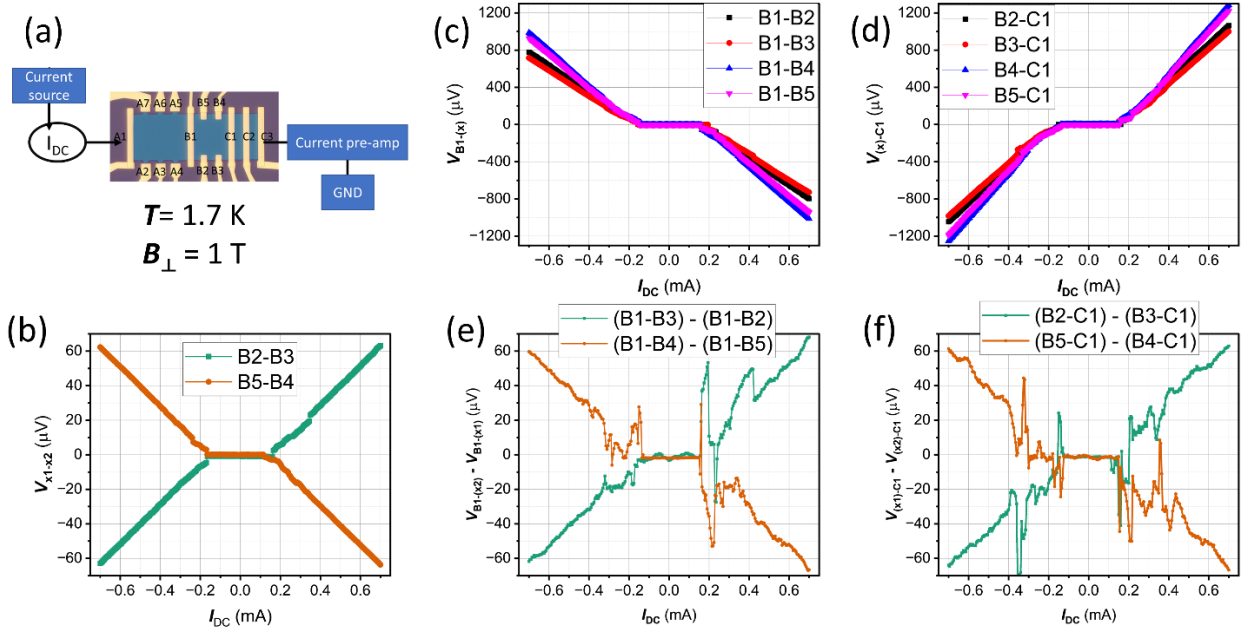

**Figure S 8**. Series resistance analysis. (a) Optical image of device. The legend in plots (b) to (d) corresponds to the contact pair used for measuring the differential DC longitudinal voltage connections. The series resistance analysis for the data from (c) is shown in (e). This is directly comparable to (b). Similarly, the analysis for the data in (d) is shown in (f) and is also comparable to (b).

Similarly, in **Figure S 8 (d)** we repeat this measurement by varying the voltage-high connection from B2 to B5 and fixing voltage-low connection on C1. Here positive slopes are seen in the normal region for all voltages and $V_{B2-C1} > V_{B3-C1}$ and $V_{B4-C1} > V_{B5-C1}$. The series resistance analysis discussed above is applied to data in **Figure S 8 (c)** and **(d)** and the corresponding results are shown in **Figure S 8Figure S 8 (e)** and **(f)**. These are both comparable to the data in **Figure S 8 (b)**, further supporting the data in the main text. This test eliminates contributions from multiple connections and wiring errors and indicates the existence of a local negative electric potential in the region of invasive contacts.

3. Hall Measurements

Here we show the Hall measurements taken on the 3-in-1 device below 80 K with a constant current of +0.3 mA injected between A1 to C3. The transverse resistance is measured in the invasive region using contacts B3-B4 (**Figure S 9 (a)-**top), B2-B5 (**Figure S 9 (a)-**middle) and as well as at the non-invasive region with A4-A5 (**Figure S 9 (a)-**bottom). The well-defined geometry of the device alleviates longitudinal contributions that can bleed into the transverse component and cause abrupt jumps seen in the Hall data on the raw flakes (**Fig. 2 (c)** of main text). The non-invasive contacts (**Figure S 9 (a)-**bottom panel) do not show linear changes in resistance with magnetic field, which is consistent with the lower signals seen in longitudinal voltages seen in this region (main text, **Fig. 3(b)**). The carrier densities (calculated using **Eq. S1**) in the invasive Hall bar region shows $R_{B2-B5}$ (**Figure S 9 (b)-**bottom) is larger than $R_{B3-B4}$ (**Figure S 9 (b)**-top). This could be because the B2-B5 contacts are located closer to the current injection and more carriers are available here. Whereas the dissipative region between B2-B3 (or B5-B4) hosting vortex flow and Bernoulli potentials leads to a decrease in available carriers downstream, resulting in lower carrier density by a factor of two in $R_{B3-B4}$.

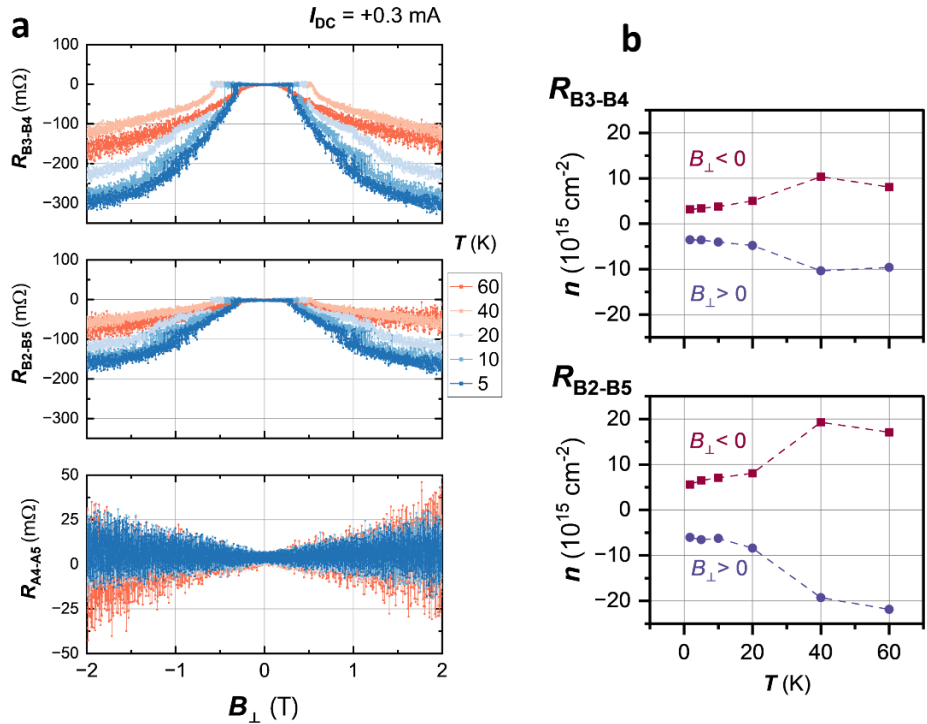

**Figure S 9**. Hall resistances on the 3-in-1 device at different temperatures below 80 K as a function of sweeping field (a) using a fixed DC current of +0.3 mA. The carrier densities for the data in top two panels is shown in (b).

**Derivation of Bernoulli potential map**

In this section, we provide the basis for the calculation of Bernoulli potentials, assuming the conductor is a type-II superconductor, and the device channel has a thickness 50 nm, width 25 µm, and length 20 µm. The schematic in **Figure S 10** aids the description that follows.

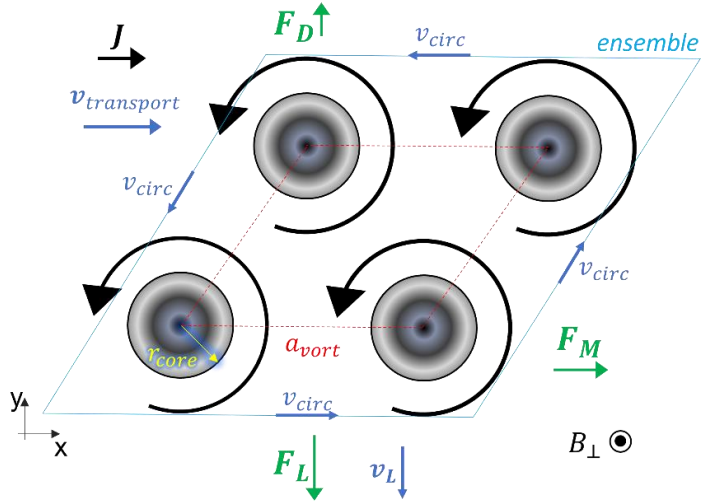

**Figure S 10**. Schematic representation of forces acting on a vortex lattice due to an applied current $J$. The vortices are grouped together and treated as an ensemble to estimate the velocity of the circulating current ($v_{circ}$) along the extremities.

The transport velocity is governed by the drift velocities of superconducting Cooper pairs and excited Bogoliubov quasiparticles. Additionally, because we are looking at the dissipative state we must include the pair-breaking (or gap) velocity. For a transport current density $J$ below the critical current density $J_c$, the superconducting condensate has a drift velocity $v_s = \frac{J(J<J_c)}{n_s e}$. And when $J > J_c$, the Bogoliubov quasiparticles have a drift velocity $v_q = \frac{J(J>J_c)}{n_r e}$. As discussed in the Hall analysis, $n_r \approx 10\%$ is the ratio of carrier densities at low temperatures ($n_s = 2 \times 10^{15}$ cm$^{-2}$, **Fig. 2d** of main text) and at 100 K ($n = 19 \times 10^{15}$ cm$^{-2}$, supplement S6). From the experimental conditions above, for a charge current I $< |\pm 0.5|$ mA, the drift velocities of the Cooper pairs and Bogoliubov quasiparticles will be minimal, $v_s < 8.3$ m/s and $v_q = 0.8$ m/s, respectively. However, there is another velocity we need to consider in the dissipative state. The gap velocity ($v_{gap}$) [3], associated with the breaking of a Cooper pair is much greater than the drift velocities, with $v_{gap} = \frac{\Delta}{k_f} = 1.76 \times 10^4$ m/s, where $\Delta = 40$ meV is the pseudogap energy, $k_f = 2m_e v_f$ is the Fermi momentum, and $v_f \approx 2 \times 10^5$ m/s is the Fermi velocity of the constituents in a Cooper pair. Therefore, due to significant dissipative transport, the gap velocity dominates. The total transport velocity is approximately $v_{transport} = v_s + v_q + v_{gap} \approx v_{gap}$.

In the mixed state, several pancake Abrikosov vortices [4] nucleate in the BSCCO sample pinned with force $F_p$. Each vortex permits a single flux quantum vector $\boldsymbol{\phi_0} = \frac{h}{2q}\frac{\boldsymbol{B}}{|\boldsymbol{B}|}$ of magnitude $2.067 \times 10^{-15}$ Wb along $\boldsymbol{B}$. As the external field increases above the Meissner state, the vortex density increases linearly $n_{vort} = \frac{B}{\phi_0}$ and the vortex lattice spacing is approximately $a_{vort} \approx \frac{1}{\sqrt{n_{vort}}}$ [5]. At $B_\perp = 1$ T, the estimated $n_{vort} = 4.8 \times 10^{10}$ cm$^{-2}$ and $a_{vort} \approx 45$ nm. The vortex circulating velocity can be estimated by $v_{circ} = \frac{\hbar}{2m^*r}$, where $r$ is the radius from the vortex core. In the vortex lattice at 1 T, the average circulating velocity of an ensemble of vortices ($r = a_{vort}/2$) is $v_{circ} \approx 637$ m/s.

The DC current through the sample induces a Lorentz force $\boldsymbol{F_L} = \boldsymbol{J} \times \boldsymbol{\phi_0}$ on the pancake vortices, causing them to drift perpendicular to the transport current when $\boldsymbol{F_L} \geq \boldsymbol{F_p}$, where $F_p$ is the pinning force on the vortices. Vortices drifting with a velocity $\boldsymbol{v_L}$, will encounter a viscous drag $\boldsymbol{F_D} = -\eta \boldsymbol{v_L} = -(\boldsymbol{F_L} - \boldsymbol{F_p})$ [6], where $\eta$ is the viscosity coefficient. It is related to the second critical field $B_{c2} = 100$ T and normal state resistivity ($\rho_n \approx 160$ μΩ cm) by, $\eta = \frac{|\phi_0|B_{c2}}{\rho_n} = 1.26e - 7$ Pa s. Anti-clockwise vortices moving towards $-y$ experience a Magnus force $\boldsymbol{F_M} = -n_s q(\boldsymbol{v_L} \times \boldsymbol{\phi_0})$ which results in a total electric field $\boldsymbol{E} = -\boldsymbol{v_L} \times \boldsymbol{B} = -\Delta\Phi$ along the longitudinal direction. Reversing $\boldsymbol{B}$, also reverses $\boldsymbol{v_L}$ and thus the $\boldsymbol{E}$ is unchanged. This is a basic reason behind why the measured longitudinal voltages do not change sign when the magnetic field is reversed. However, to capture the experimental result ($V_{xx,2} \approx -V_{xx,1}$) we need to implement the circulation velocity of the vortex ensemble and modulations due to device geometry to calculate the Bernoulli potential and corresponding electric field.

The potential difference ($\Phi$) created by the Magnus force gives rise to the Bernoulli potential we measure by solving for the independent velocities at the top and bottom edges of the device. In the mixed state at high currents, vortices move from $+y$ edge to the $-y$ with a velocity ($v_L = 5.7$ m/s). The stream of vortices encounter accumulated vortices along the bottom edge and enter the contacts at a reduced velocity ($-v_L/2$). The vortex bunching around the periodic indentations created by invasive contacts create a sinusoidal modulation with an amplitude (A) and phase $2\pi x/L$, where $L$ is the device length. The amplitude here is set to be the ratio of the total contact overlap width ($2 \times 5$ μm) over the device width ($W = 25$ μm). Lastly, the contribution from the Magnus force acting along the longitudinal direction leads to factors $(1 - v_L)$ and $(1 + \frac{v_L}{2})$ with respect to the circulation velocities at the top and bottom respectively. All these modifications result in the top and bottom velocity terms:

$$v_{top}(x) = v_{circ}\left(1 + A\sin\left(\frac{2\pi x}{L}\right)\right)(1 - v_L)$$

$$v_{bot}(x) = v_{circ}\left(1 + A\sin\left(\frac{2\pi x}{L}\right)\right)\left(1 + \frac{v_L}{2}\right)$$

**Eq. S1**

The longitudinal velocities are multiplied by corresponding spatial scaling factors $w_{top} = 1\ to\ 0$ and $w_{bot} = 1 - w_{top}$ to account for the reduction in magnitude of vortex velocities in the middle of the ensemble. The resulting longitudinal component of vortex velocity is

$$v_{vortex}(x) = v_{top}(x)w_{top} + v_{bot}(x)w_{bot} \qquad \textbf{Eq. S2}$$

Since the transport velocities are also along the longitudinal direction, the net longitudinal velocities are $\sum v(x) = v_{transport}(x) + v_{vortex}(x)$. The transverse component of vortex velocity is mainly dictated by the Lorentz component, so $v(y) = v_L$. The sum of squares of the velocity term, $\left(\sum v(x,y)\right)^2 = v(x)^2 + v(y)^2$, is used in the calculation of the Bernoulli potential (main text **Fig. 4(b)**).